\documentclass[letterpaper,english,aps,prb,twocolumn,floatfix,showpacs,amsfonts,amssymb,superscriptaddress]{revtex4}
\usepackage[T1]{fontenc}
\usepackage[latin1]{inputenc}
\usepackage{graphics}
\usepackage{amssymb}
\usepackage{amsmath}
\usepackage{overpic}

\newcommand{\be}{\begin{equation}}
\newcommand{\ee}{\end{equation}}
\newcommand{\bea}{\begin{eqnarray}}
\newcommand{\eea}{\end{eqnarray}}

\newcommand{\rmd}{{\rm d}}

\newcommand{\cuoo}{CuO$_{2}$}
\newcommand{\lco}{La$_{2}$CuO$_{4}$}
\newcommand{\lcoo}{La$_{2}$CuO$_{4+y}$}
\newcommand{\lasco}{La$_{2-x}$Sr$_{x}$CuO$_{4}$}

\renewcommand\d{\delta}

\renewcommand\t{\tau}

\newcommand\s{\sigma}

\newcommand\D{\Delta}

\newcommand\ra{\rightarrow}

\newcommand\pll{\parallel}

\newcommand\pd{\partial}

\newcommand\bH{{\bf H}}
\newcommand\bD{{\bf D}}

\newcommand\bn{{\bf n}}
\newcommand\bx{{\bf x}}

\newcommand\cA{{\cal A}}

\newcommand\nn{\nonumber}
\newcommand\lb{\label}
\def\pref#1{(\ref{#1})}

\newcount\bozza \bozza=0
\ifnum\bozza=1
\newdimen\shift \shift=-2truecm
\def\lb#1{%
{\label{#1}\rlap{\kern\shift{$\scriptstyle#1$}}}}
\else\def\lb#1{\label{#1}} \fi

\begin{document}

\title{Impurity susceptibility and the fate of spin-flop transitions in 
lightly-doped {\lco}}

\author{M.~B.~Silva~Neto}

\email{barbosa@itp3.uni-stuttgart.de}

\affiliation{Institut f\"ur Theoretische Physik, Universit\"at
Stuttgart, Pfaffenwaldring 57, 70550, Stuttgart, Germany}

\author{L.~Benfatto}

\email{lara.benfatto@roma1.infn.it}

\affiliation{Centro Studi e Ricerche ``Enrico Fermi'', via Panisperna 89/A,
00184, Rome Italy}
\affiliation{CNR-INFM and Department of Physics, University of Rome ``La
  Sapienza'',\\ Piazzale Aldo Moro 5, 00185, Rome, Italy}

\date{\today}

\begin{abstract}

We investigate the occurrence of a two-step spin-flop transition and spin
reorientation when a longitudinal magnetic field is applied to lightly
hole-doped {\lco}. We find that for large and strongly frustrating
impurities, such as Sr in {\lasco}, the huge enhancement of the
longitudinal susceptibility suppresses the intermediate flop and the
reorientation of spins is smooth and continuous. Contrary, for small and
weakly frustrating impurities, such as O in {\lcoo}, a discontinuous spin
reorientation (two-step spin-flop transition) takes place. Furthermore, we
show that for {\lasco} the field dependence of the magnon gaps differs
qualitatively from the {\lco} case, a prediction to be verified with
Raman spectroscopy or neutron scattering.

\end{abstract}

\pacs{74.25.Ha, 75.10.Jm, 75.30.Cr}

\maketitle

{\it Introduction }$-$ Besides being the parent compound of
high-temperature superconductors, undoped {\lco} (LCO) exhibits
remarkable and unusual magnetic properties that have received a great deal
of attention in the past few years.  These properties stem mostly from the
combination of low crystal symmetry (in the low temperature orthorhombic
phase) and spin orbit coupling that allows for the appearance of
Dzyaloshinskii-Moriya (DM) interactions and result in the occurrence of
phenomena such as: weak ferromagnetism,\cite{Thio-I} anisotropic magnetic
response,\cite{Lavrov} field-induced spin reorientation,
\cite{Gozar,Reehuis} and spin-flop transitions,\cite{Thio-I,Reehuis} among
others. These many aspects of such unconventional antiferromagnetic
material have been thoroughly explored experimentally with Raman
spectroscopy, neutron scattering, and magnetic susceptibility measurements,
and are at present fully understood from the theoretical point of view, the
agreement between theory and experiment being
remarkable.\cite{Benfatto,MLVC,Luscher-Sushkov}

When few holes are introduced into {\lco}, the long-range antiferromagnetic
order is rapidly destroyed, for example at $x\approx 0.02$ in {\lasco}
(LSCO). The doped holes, which are trapped by the strong ionic potential
from the dopants, induce a local spin distortion which {\it frustrates}
the antiferromagnetic interactions and eventually leads to the complete
suppression of the antiferromagnetism. The amount of frustration introduced
through doping depends crucially on two aspects: i) the strength of the
ionic trap potential provided by the shallow acceptor; ii) the spatial
position the dopant goes inside the crystal. It has been shown that for Sr
acceptors,\cite{Luscher} which are located (out of the plane) at the center of the Cu
plaquettes and provide a weaker potential, frustration is maximized, while
for O dopants, which enter interstitially into the matrix and provide a
stronger ionic potential, frustration is expected to be much
smaller.\cite{Sushkov-private} This scenario is consistent with the fact
that the N\'eel temperature is suppressed much more rapidly for Sr dopants
than for O ones,\cite{Lavrov} and it is also consistent with recent
magnetic-susceptibility measurements which show a large impurity
contribution to the longitudinal susceptibility (a direct measure of
frustration, as we shall see below) for {\lasco}, while this is negligible
for {\lcoo} (LCOy).\cite{Lavrov}

\begin{figure}[t]
\includegraphics[scale=0.23]{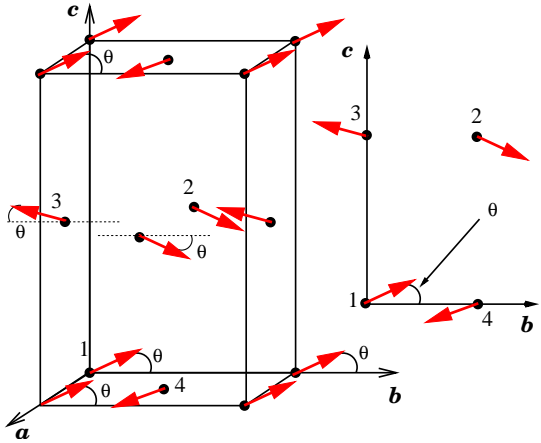}
\caption{(Color online): Magnetic structure of {\lco} for a small
longitudinal field, $H\parallel b$. Solid (red) arrows represent
Cu$^{++}$ moments and for $H=0$ we have $\theta=0$.}
\label{Fig-LCO}
\end{figure}

The natural question to be answered now is: {\it how are the magnetic
phenomena of {\lco} listed above affected by frustration upon doping?} In
what follows we will focus on the fate of the spin reorientation and
spin-flop transitions when a magnetic field is applied along the in-plane
orthorhombic $b$ (longitudinal) direction to Sr- or O- doped {\lco}. In the
presence of a longitudinal field the DM interaction causes the Cu$^{++}$
spins, initially oriented along $b$ at zero field (see Fig.\ \ref{Fig-LCO}
at $\theta=0$), to gradually develop an out-of-plane component, which fully
orients the spins along the $c$ direction above a certain critical field
$H_c^2$.\cite{Benfatto,Luscher-Sushkov,Thio-II,CP} Moreover, the longitudinal 
field is expected to cause a spin-flop when $H$ equals the smaller of the 
transverse gaps. In the case of undoped LCO this means that at an intermediate
field $H_c^1<H_c^2$, of order of the in-plane DM gap, a spin-flop
transition of the in-plane spin component is 
expected,\cite{Benfatto,Luscher-Sushkov,Thio-II,CP} with
the spins aligning in the $ac$ plane.  Even though the rotation angle
$\theta$ is continuous at the transition, its field dependence (slope)
changes, giving rise to a kink in the $\theta(H)$ curve. The issue is
whether this intermediate flop is actually present in doped LCO.

A very important clue to the answer for this question comes from
magnetoresistance (MR) experiments. Indeed, as it has been shown recently in
Ref. \onlinecite{mrpaper}, the spin reorientation for longitudinal fields
causes an increase of the localization length of the trapped carriers,
which enhances their hopping conductivity and leads to a large negative
MR. It turns out that the relative MR is a direct
measurement of the field dependence of the angle $\theta(H)$.\cite{mrpaper}
Thus, the kink of $\theta(H)$ at $H_c^1$ should leave an imprint in the MR
curves. However, different $\theta(H)$ behaviors have been obtained for
different types of acceptors (O or Sr).  While the early data from
Thio {\it et al.}  \cite{Thio-II} {\it clearly indicate} that such an
intermediate SF transition indeed occurs in O-doped (LCOy), and it
is manifest as a {\it kink} in the MR curves, the very recent MR
experiments by Ono {\it et al.} in untwinned LSCO single crystals have
shown {\it no sign whatsoever} of an intermediate SF transition.\cite{Ono}
As we shall now explain, the suppression of the intermediate flop in
LSCO is a direct consequence of the strongly frustrating character of
the Sr acceptors. In addition, we show that the field dependence of the
magnon gaps in the doped case can be qualitatively different from the
undoped case depending on the amount of frustration introduced by doping, 
a prediction which can be tested by means of one-magnon Raman spectroscopy 
or Neutron scattering.

%
%
%

{\it The model }$-$ We start with a non-linear sigma model description for 
the low-energy dynamics of the spin degrees of freedom in undoped {\lco},\cite{Benfatto} 
which incorporates the DM and XY anisotropies ($\beta=1/T$ and
$\int=\int_{0}^{\beta}\rmd\tau\int\rmd^{2}{\bf x}$)
\bea
\lb{nlsm}
{\mathcal S}_{n}=\frac{\chi_\perp}{2}\sum_m
\int \left\{(\partial_{\tau}{\bf n}_m)^{2}+
  c^2(\nabla{\bf n}_m)^{2}+\right. \nn\\
\left. ( \D_{in} {n}_m^a)^2+( \D_{out} {n}_m^c)^2+
\eta(\bn_m-\bn_{m+1})^2
\right\}.
\eea
Here $\bn_m$ is a continuous unit-length vector field which represents the
three components of the staggered magnetization in the $m^{th}$ plane along
the $(a,b,c)$ orthorhombic directions, $\chi_\perp$ is the transverse
susceptibility, $c$ the spin-wave velocity, $ \eta=2JJ_\perp$ (with
$J,J_\perp$ in-plane and out-of plane superexchange respectively), and
$\D_{in}$ ($\D_{out}$) is the in-plane (out of plane) gap, whose value is
controlled by the DM (XY) anisotropy. At zero magnetic field the
ground-state of the action \pref{nlsm} is given by $\bn_m=\sigma_0
\hat\bx_b$, where $\hat\bx_b$ is the unit vector in the $b$ direction, and
$\sigma_0\leq 1$ is the order parameter renormalized by both quantum and
thermal fluctuations. There is almost perfect antiferromagnetic (AF) N\'eel
order within each {\cuoo} layer (up to a tiny canting staggered along the
$c$ axis due to DM interactions, not shown in Fig.\ 1), while spins in
neighboring layers exhibit AF and ferromagnetic order along the $ac$ and
$bc$ planes, respectively (see Refs.\ \onlinecite{Benfatto,Reehuis} and
references therein). At finite magnetic field the following terms should be
added to the action \pref{nlsm}\cite{Benfatto}
\bea
{\cal S}_{nH}=\frac{\chi_\perp}{2}\sum_m\int \left[ 2i
\bH\cdot (\bn_m \times \pd_\t \bn_m)-\bH^2\right.\nn\\
\lb{nlsmh}
\left.+(\bH\cdot \bn_m)^2
-(-1)^m2\bH\cdot (\bn_m\times \bD)\right],
\eea
where $\bD=D\hat\bx_a$ is the DM vector and we measured the magnetic field
in units of $g_s\mu_B$, where $g_s\approx 2$ is the gyromagnetic ratio and
$\mu_B$ is the Bohr magneton.  For $\bH\pll b$ this last term can be
written as $(-1)^mHDn_m^c$, and it is responsible for the development of a
finite $n_m^c$ component of the order-parameter, i.e. to a continuous
rotation of the spins in the $bc$ plane with $\langle
\bn_m\rangle=(0,\s_0\cos\theta,(-1)^m\s_0\sin\theta)$, where $\theta$ is
the angle the spins form with the $ab$ plane, see Fig.\ \ref{Fig-LCO}. By
adding transverse fluctuations to $\langle \bn_m\rangle$ one can compute
the value of the in-plane and out-of-plane gap as a function of magnetic
field.\cite{Benfatto} One then finds that the in-plane gap (i.e. the gap
for the $a$ fluctuations) decreases, and vanishes at a critical field
$H_c^1=\D_{in}$.\cite{Benfatto,Luscher-Sushkov,Thio-II,CP}  As a consequence, 
at $H=H_c^1$ the spins perform an in-plane spin-flop, $\langle \bn_m\rangle
=(\s_0\cos\theta,0,(-1)^m\s_0\sin\theta)$, and orient in the $ac$ plane.
Although $\sin\theta$ is a continuous function of the field across $H_c^1$,
its slope changes (see left panel of Fig.\ 2)
\bea 
\lb{h<h1}
\sin\theta= 
    \frac{H D/\sigma_0}{\Delta_{out}^2+4\eta-H^2}, \quad
    0<H<H_c^1\\
\lb{h>h1}
\sin\theta= 
    \frac{H D/\sigma_0}{\Delta_{out}^2+ 4\eta-\Delta_{in}^2},
    \quad H_c^1<H<H_c^2
\eea
leading to a kink in the field dependence of $\theta(H)$.\cite{Benfatto} 
At $H\geq H_c^2$ the spins are fully oriented along $c$ ($\sin\theta=1$). 

\begin{figure}[htb]
\includegraphics[scale=0.33,angle=-90]{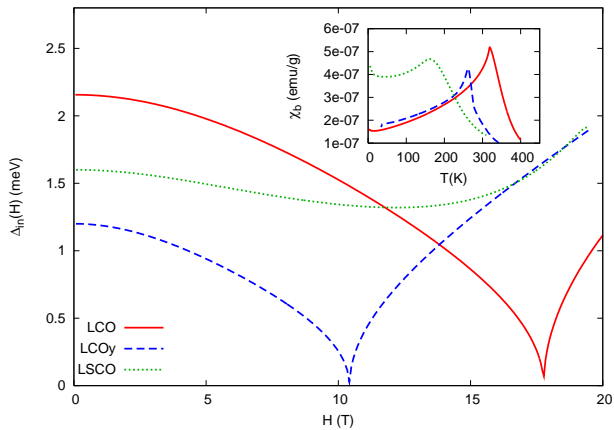}
\caption{(Color online): Field dependence of the in-plane (DM) gap for
  undoped LCO and for doped LCOy and LSCO. In the first two cases
  $\chi_{imp}=0$ and $\chi_{imp}=5$, so that the low-field susceptibility
  has approximately the same value (see inset), and an intermediate spin
  flop occurs at $H_c^1$ given by Eq.\ \pref{Bc1}, where the gap
  vanishes. Instead, for doped LSCO $\chi_{imp}=120$, $\chi_b$ is strongly
  enhanced (see inset), no intermediate spin flop occurs, and the
  in-plane gap never vanishes. Inset: low-field susceptibility data taken
  from Ref. [\onlinecite{Lavrov}].}
\label{Fig-gaps}
\end{figure}

{\it Longitudinal spin susceptibility} $-$ The possibility to observe the same
feature at finite doping depends crucially on the type of acceptor, O or
Sr, introduced in host {\lco}. At low doping the holes are localized by the
Coulomb trap potential provided by the dopants.  The hole wave function is
given by $\psi(\bx)=\Psi \chi(\bx)$, where $\Psi$ is a two-component spinor
accounting for the pseudospin degeneracy (the hole can reside in either up
or down sublattices), and $\chi(\bx)\sim e^{-\kappa x}$ is an hydrogen-like
localized state with inverse localization length $\kappa$, describing the
spatial dependence of the wave function. The coupling between the holes
pseudospin ${\bf d}=\Psi^\dag\sigma\Psi$ and the background magnetization
leads to a {\it partial frustration} of the AF order, i.e. to a (local)
spiral distortion of the N\'eel phase and to a softening of the magnon gaps
(at $H=0$) with respect to the undoped case.\cite{MVC,Luscher} Besides this
local effect, it has been proposed in Ref.\ [\onlinecite{Luscher}] that in
the presence of a longitudinal magnetic field (i.e. $\bH\pll b$) a new
global Zeeman coupling between the holes pseudospin and the magnetic field
is present
\be
\lb{snh}
{\cal S}_{H\psi}=-\frac{\delta}{2}\sum_m \left[ d_\pll H (n_m^b)^2\right],
\ee
where $\delta$ is the doping and $d_\pll$ is the field-induced pseudospin
component along the field direction that can be quite generically expressed 
as
\be
\lb{dpll}
d_\pll=
\chi_\perp\chi_{imp} H.
\ee
The exact value of $\chi_{imp}$ depends on the microscopic details of the
problem, such as the strength of the trap potential and the spatial
distribution of the dopants.\cite{Luscher}
In what follows, however, we shall assume that $\chi_{imp}$ is a 
phenomenological parameter that can be directly extracted from the 
enhancement of the longitudinal susceptibility $\chi_b$ upon doping. 
Indeed, from the action \pref{nlsm} (for $\bH\pll b$) and \pref{snh} 
one can easily derive the spin susceptibility along $b$ in linear-response 
theory as $\chi_b=(1/\beta V)\partial^{2}\log{Z}/\partial 
H^{2}|_{H=0}$,\cite{MLVC} where $Z({\bf H})=\int{\cal D}{\bf n}\exp
\left\{-\left({\cal S}_n+{\cal S}_{nH}+{\cal S}_{H\psi}\right)\right\}$ 
is the Euclidean partition function for the total action.
The result is
\be
\lb{susc}
\chi_b=\chi_{b}^{u}+\frac{\chi_\perp D^2}{\D_{out}^2+4\eta}+
\chi_\perp\d\chi_{imp},
\ee
where
$\chi_{b}^{u}=\chi_\perp[\langle (n_m^a)^2 + (n_m^c)^2 \rangle-4\langle
n_m^a\pd_\tau n_m^c \rangle^2]$.\cite{MLVC}  In a conventional (non DM) AF only the
first term in Eq.\ \pref{susc} contributes, and since $\chi_u^b$ vanishes
at $T=0$ one recovers the expected vanishing of the longitudinal
susceptibility. In undoped {\lco} the DM interaction leads to the second
term of Eq.\ \pref{susc}, and then to a finite longitudinal response even
at $T=0$.\cite{MLVC} When the system is doped, the trapped holes (impurities)
contribute to $\chi_b$ with the last term in Eq.\ \pref{susc}, leading to
an even larger positive increase of the longitudinal susceptibility
proportional to $\chi_{imp}$.\cite{Luscher}  From the measurements of Ref.\
\onlinecite{Lavrov}, shown in the inset of Fig.\ \ref{Fig-gaps}, we see
that while doping with O changes only slightly $\chi_b$, leading to
$\chi_{imp}\sim {\cal O}(1)$, doping with Sr leads to a longitudinal
susceptibility four times larger than in the undoped case, leading to
$\chi_{imp}\sim 100$.

%
%
\begin{figure}[t]
\begin{center}
\includegraphics[scale=0.33,angle=-90]{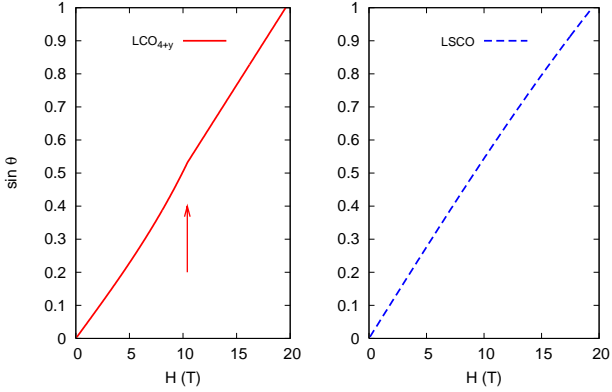}
\caption{(Color online): Field dependence of the rotation angle $\theta(H)$ 
for $1\%$ doped LCOy (left) and LSCO (right). Observe that for LCOy a kink at
the intermediate spin-flop transition at $H_c^1$ is observed, as indicated 
by the arrow.}
\end{center}
\label{Fig-angle}
\end{figure}
%

{\it The two-step spin-flop transition} $-$ To investigate the effect of
hole doping on the spin-flop transitions we calculate the field dependence
of the magnon gaps in the presence of the impurity contribution \pref{snh}.
Here we follow the same procedure described in Ref.\
\onlinecite{Benfatto,Luscher-Sushkov} for undoped {\lco}, by simply replacing in
Eq.\ \pref{nlsmh}
\be
\lb{rep}
H^2(n_m^b)^2 \ra H^2(1-\delta\chi_{imp})(n_m^b)^2.
\ee
As a consequence, the field evolution of the the out-of-plane canting angle 
$\theta$ is given by
\be 
\lb{Eq-Theta} 
\sin\theta= \frac{H D/\sigma_0}
{\Delta_{out}^2(\delta)+4\eta-(1-\delta\;\chi^b_{imp})H^2},
\ee
instead of Eq.\ \pref{h<h1}, valid for the undoped system. To account for the 
rotation of the order parameter with the field, we introduce fluctuations 
$n_m^a,n_m^{c'}$ orthogonal to the ground-state configuration $\langle 
\bn_m\rangle$ as $\bn_m=\langle 
\bn_m\rangle +(n_m^a, \s_0\cos\theta-(-1)^m\sin\theta
n_m^{c'}, (-1)^m\s_0\sin\theta+\cos\theta n_m^{c'})$. 
The spectral function of each fluctuating mode has a 
two-peak structure,\cite{Benfatto} given by
\be 
\lb{ca} 
\cA^{a,c}(\omega>0)=
\left[Z^{a,c}_+\d(\omega-\omega_+)+
Z^{a,c}_-\d(\omega-\omega_-)\right],
\ee
where $\omega_\pm$ are the eigenvalues of the matrix of the transverse
fluctuations
\bea
\omega_\pm^2&=&\frac{x_1^2+x_2^2+4H^2\cos^2\theta}{2}\pm \nn\\
\lb{opm}
&+&\frac{1}{2}\sqrt{(x_1^2+x_2^2+4H^2\cos^2\theta)^2-4x_1^2x_2^2},
\eea
and we defined
\bea
x_1^2&=&\D_{in}^2-H^2(1-\d\chi_{imp}),\nn\\
x_2^2&=&[\D_{out}^2-H^2(1-\d\chi_{imp})]\cos^2\theta+2\eta(1-\cos(2\theta)).\nn
\eea
The spectral weights 
\bea
Z^{a}_\pm= \mp({-\omega_{\pm}^2+x_{2}^2})/ 2({\omega_+^2-\omega_-^2})\omega_\pm,\nn\\ 
Z^{c}_\pm= \mp \cos^2\theta
({-\omega_{\pm}^2+x_{1}^2})/ 2({\omega_+^2-\omega_-^2})\omega_\pm,\nn
\eea
allow one to identify the leading pole for each mode. For example, for
$H\ra 0$ we have $Z^a_-\gg Z^a_+$, so that $\omega_-$ identifies the
evolution of the in-plane (or DM) gap at small field, and $\omega_+$
identifies the out-of-plane (or XY) gap, while as $\theta\ra \pi/2$ the
situation is reversed.\cite{next} From Eqs.\ \pref{opm}, using $\D_{in}<\D_{out}$, one
sees that the in-plane gap (given by the $\omega_-$ solution) vanishes when
$x_1^2=0$, i.e. at the critical field
\be
\lb{Bc1} 
H_{c}^1(\delta)=
\frac{\Delta_{in}}{\sqrt{1-\delta\,\chi^b_{imp}}}.
\ee
Above $H_c^1$ a spin-flop occurs, the spins rotate in the $ac$ plane with
the angle $\theta$ described by Eq.\ \pref{h>h1}, and the gaps evolve
according to
\bea
\lb{deltain}
\omega_{in}^2
&=&H^2(1-\d\chi_{imp})-\D_{in}^2,\\
\omega_{out}^2&=&(\D_{out}^2-\D_{in}^2)
\cos^2\theta+2\eta(1-\cos(2\theta)).
\eea
%

%
%
%

Eq.\ \pref{Bc1} is the central result of our paper. It tells us that the
larger the frustration introduced with doping, $\delta\chi_{imp}\ra 1$, the
larger $H_c^1$ will become (with no solution at all for
$\delta\chi_{imp}>1$). Eventually the first critical field $H_c^1$ becomes
larger than the second critical field $H_c^2$, at which spins are fully
polarized along $c$, and as a consequence no intermediate spin-flop
occurs. Consistently, one would expect that in this case the in-plane gap
does not vanish, as it follows indeed from the gaps equation
\pref{opm}. For {\lco} all the parameter values are extracted from Raman
experiments:\cite{Benfatto} $D=\Delta_{in}=2.16$ meV, $\Delta_{out}=4.3$
meV, $\eta=1 (meV)^2$ and $\sigma_0=0.5$. At finite doping $D$ and $\eta$
are almost unchanged, while one expects a softening of the gaps due to the
hole doping.\cite{MVC,Luscher} Finally, $\chi_{imp}$ is extracted, according to
Eq.\ \pref{susc}, from the low-field susceptibility data,\cite{Lavrov} and
using the values of $H_c^2$ measured by
magnetoresistance\cite{Thio-II,Ono} one can also estimate $\sigma_0$,
which enters in the field dependence \pref{h<h1}-\pref{Eq-Theta} of the
canting angle $\theta$. For {\lasco} at $x=0.01$ we have $\D_{in}=1.55$
meV\cite{Gozar}, $\D_{out}=3.2$ meV\cite{MVC}, $\sigma_0=0.32$, and
$\chi_{imp}=120$. Such large value of $\chi_{imp}$ implies that no
intermediate spin flop occurs, $\theta$ increases smoothly with the applied
field according to Eq.\ \pref{Eq-Theta}, as shown in the right panel of
Fig.\ 2, and no features are expected in 
the MR curves.\cite{Ono,mrpaper} Furthermore, the in-plane gap softens only
slightly with the field but never vanishes.  This signals the strongly
frustrating character of the Sr dopants. Moreover, a spectral-weight
redistribution between the two poles of the spectral function \pref{ca} is
expected, that will be discussed elsewhere.\cite{next} For {\lcoo} at
$y=0.01$ one has\cite{Thio-II} $\D_{in}=1.2$ meV, $D=1.6$ meV,
$\Delta_{out}=2.6$ meV, $\sigma_0=0.4$, and
$\chi_{imp}=5$. The impurity contribution to the low-field susceptibility
is negligible (see inset Fig.\ \ref{Fig-gaps}), and the first critical
field \pref{Bc1} is just slightly larger than the value from $\D_{in}$,
around $\approx 10$ T.\cite{Thio-II} Thus, $\theta$ increases according to Eq.\
\pref{h<h1}, with a kink at $H_c^1$ that shows up as a knee in the
magnetoresistance,\cite{Thio-II} which is proportional to
$\sin^2\theta$.\cite{mrpaper} At the same time the in-plane gap softens
with increasing field, it vanishes at $H_c^1$ and increases again at larger
field, according to Eq.\ \pref{deltain}, following the same behavior
measured in the undoped compound\cite{Gozar,Benfatto}.

{\it Conclusions} $-$ We have investigated the influence of frustration on
the sequence of spin-flop transitions in lightly hole doped {\lco}. We have
demonstrated that for strongly frustrating dopants, which have a large
impurity susceptibility and give rise to a large $T=0$ longitudinal 
susceptibility (a {\it direct measure of frustration}), the effects of a 
longitudinal magnetic field on the underlying Cu$^{++}$ spins is weakened. 
As a result, the in-plane gap depends only softly on the applied field and 
never vanishes. Thus, while for weakly frustrating impurities, like in 
{\lcoo}, the intermediate SF transition is in fact present, it is completely 
suppressed for strongly frustrating impurities, like in {\lasco}. Finally, we 
predict that for {\lasco} the magnetic field dependence of the magnon gaps 
{\it differs qualitatively} from the observed behavior in undoped 
{\lco}\cite{Gozar,Benfatto} and thus we propose one-magnon Raman spectroscopy 
or neutron scattering as {\it smoking gun} experiments to be performed in 
order to give support to the underlying mechanism of trapped-holes inducing 
local spiral distortions.\cite{Luscher}

The authors acknowledge invaluable discussions with Yoichi~Ando,
B.~Keimer, A.~Lavrov, and O.~Sushkov.

\end{document}